\documentclass[aps,prl,twocolumn,showpacs,floatfix,a4paper]{revtex4}
\usepackage{graphicx}
\usepackage[intlimits]{amsmath}

\begin{document}

\title{Two particle excited states entanglement entropy in a one-dimensional 
ring}
\author{Richard Berkovits}
\affiliation{Department of Physics, Bar-Ilan
University, Ramat-Gan 52900, Israel}

\begin{abstract}
The properties of the entanglement entropy (EE) of two particle
excited states in a one-dimensional ring are studied.
For a clean system we show analytically that as long as the momenta of
the two particles are not close, the EE is twice the value
of the EE of any single particle state. For almost identical momenta
the EE is lower than this value. The introduction of disorder is numerically
shown to lead to a decrease in the median EE of a two particle excited 
state, while interactions (which have no effect for the clean case) mitigate
the decrease. For a ring which is of the same size as the localization
length, interaction increase the EE of a typical two particle excited state
above the clean system EE value.
\end{abstract}


\maketitle

\section{Introduction}

There has been a growing interest in the behavior of entanglement entropy (EE)
\cite{amico08} in different physical fields. In condensed mater, much
of the interest stems from the behavior of the ground state EE 
in the presence of quantum phase transitions (QPTs)
\cite{amico08,vojta06,lehur08,goldstein11}. The EE of a finite region A of a 
one-dimensional system grows logarithmically as long as the region's size
$L_A$ is smaller than the correlation length $\xi$ characterizing the system,
while it saturates for $L_A>\xi$ \cite{holzhey94,vidal03,calabrese04}. 
This behavior may be used in order to extract
$\xi$, for example the ground state localization length of the Anderson 
transition \cite{berkovits12}. 

A natural question is what is the behavior of the EE for the excited states
\cite{alba09} ? Beyond the growing interest in EE coming from the quantum
information circles, the question 
whether EE is a useful concept in studying the behavior
of excited states, is relevant to the condensed matter community. 
Low lying excited states in the vicinity of 
a ground state quantum critical point (QCP) should be strongly influenced 
by the critical point \cite{schadev99}, and one expects it to show in the 
behavior of the EE of these states. Moreover, the whole concept of the 
many-body localization transition \cite{AGKL97,gornyi05,basko06}
is centered on the behavior of the excited states. 
The localization-delocalization transition occurring at a critical excitation
energy should change the properties of excitations above it which should
be manifested in the properties of the excited states.
Although much effort
went into trying to understand the transition using different properties
of the excited states (such as level statistics, inverse participation ratio, 
conductance, and correlations) 
\cite{berkovits98,oganesyan07,monthus10,berkelbach10,pal10,canovi11,cuevas12},
all these studies were performed for rather small systems, and 
many questions remain open. 
Recently \cite{bardarson12}, time evolution of the entanglement of an initial
state was studied, and it showed signs of many-particle delocalization.
Thus, EE seems as a useful tool to study the many-body localization transition.

Unlike the ground state EE for which universal results exist, the 
understanding of EE for the excited states is still a work
in progress \cite{alcaraz08,masanes09,berganza12}. Therefore, it would be
useful to consider a system for which the EE of the excited states
is simple enough to describe analytically, although it exhibits interesting 
behavior such as interaction induced delocalization of the excited states. 
In this paper we study the EE for such a system, namely two 
particles on a ring.
The study of two interacting particle (TIP) in a disordered 
one dimensional system has a long history in the context
of many particle delocalization problem. All single-electron states for any
amount of disorder are localized \cite{lee85}. This continues to be true for
two-electron states, however, the localization length becomes longer as the
repulsive interaction becomes stronger
\cite{shepelyansky94,imry95}. This interaction induced delocalization was
confirmed numerically \cite{frahm95,weinmann95,vonoppen96,jacquod97}. 
It is important to emphasize that there is no enhancement of the localization 
length in the ground state. The delocalization becomes significant only for 
higher excitations.

\section{Clean Ring}

For a clean ring composed of $N$ sites, the tight-binding Hamiltonian is
given by:
\begin{eqnarray} \label{clean_hamiltonian}
H &=& 
\displaystyle \sum_{j=1}^{N} \epsilon_j {\hat a}^{\dagger}_{j}{\hat a}_{j}
-t \displaystyle \sum_{j=1}^{N}
(e^{i \alpha} {\hat a}^{\dagger}_{j}{\hat a}_{j+1} + h.c.), 
\end{eqnarray}
where for the clean case $\epsilon_j=0$, $t=1$ is the
hopping matrix element between neighboring sites, 
and ${\hat a}_j^{\dagger}$ is the 
creation  operator of a spinless electron at site $j$ on the ring. 
In order to break symmetry (we shall see why this is important further on)
a magnetic flux $\phi$ threading the ring is introduced, 
where $\alpha=2 \pi \phi/(\phi_0 N)$,
and $\phi_0=hc/e$ is the quantum flux unit. The single-electron eigenvalues are
$\varepsilon(k)=-2t \cos(p-\alpha)$ where $p= 2 \pi k/N$, and 
$k=0,\pm 1, \pm 2,\ldots \pm N/2$. The eigenvectors are given by
\begin{eqnarray} \label{single_particle}
|k\rangle = (1/\sqrt{N})\displaystyle \sum_{j=1}^{N} 
\exp(\imath p j) {\hat a}^{\dagger}_j|\emptyset \rangle,
\end{eqnarray}
where $|\emptyset \rangle$ is the vacuum state.

The two particle eigenvalues are $\varepsilon(k1,k2)=-2t \left(\cos(p_1-\alpha)
+\cos(p_2-\alpha)\right)$. The eigenvector 
\begin{eqnarray} 
\label{two_particle}
|k_1,k_2\rangle = (1/N)\displaystyle \sum_{j_1>j_2=1}^{N} 
A(k_1,j_1,k_2,j_2) {\hat a}^{\dagger}_{j_1} {\hat a}^{\dagger}_{j_2}
|\emptyset \rangle, 
\end{eqnarray}
where
\begin{multline}\label{aa}
A(k_1,j_1,k_2,j_2) =  \\
\big( \exp(\imath (p_1 j_1 +p_2 j_2)) 
- \exp(\imath (p_1 j_2 +p_2 j_1)) \big).
\end{multline}

Once the eigenvectors of the system are available one can (in principal) 
calculate the
EE. The entanglement between a region A (of length $N_A$)
in the system and the rest of
the system (denoted by B) for a given eigenstate $|\Psi\rangle$
is measured by the EE $S_{A/B}$. 
This EE is related to the region's reduced density matrix $\rho_{A/B}$, defined 
in the following way: 
\begin{eqnarray} \label{density_matrix}
\hat \rho_{A/B}={\rm Tr}_{B/A}|\Psi\rangle\langle \Psi |, 
\end{eqnarray}
where the trace is over region's B or A degrees of freedom. The EE is
related to the eigenvalues $\lambda_i$ of the reduced density matrix: 
\begin{eqnarray} \label{entropy}
S_{A/B}=-\Sigma_{i}\lambda_{i}\ln(\lambda_i).
\end{eqnarray}
One important result of this definition is the symmetry between the EE
of the two regions $S_A=S_B$. 

Following Cheong and Henley \cite{cheong04}, one can write 
the pure state $|\Psi\rangle=
\displaystyle \sum_i|i_A\rangle|\phi_{B,i}\rangle$,
where $|i_A\rangle$ is a complete orthonormal many-body basis of region A,
while $|\phi_{B,i}\rangle$ is the state in region B associated with  
$|i_A\rangle$. Please note that $|\phi_{B,i}\rangle$ is not normalized. Using
that notation, the reduced density matrix,
\begin{eqnarray} \label{reduced_density_matrix}
\hat \rho_{A}=\displaystyle \sum_{i,j=1}^{N}|i_A\rangle\langle j_A|, 
\end{eqnarray}
or in matrix form:
\begin{eqnarray} \label{reduced_density_matrix_1}
\rho_{A}(i,j)=\langle\phi_{B,i}|\phi_{B,j}\rangle.
\end{eqnarray}
Utilizing the occupation basis in the A region, i.e., 
$|i_A\rangle=|n^i_1,n^i_2,n^i_3,\ldots,n^i_{N_A}\rangle$ (where $n^i_j=0,1$), 
one can define an operator, ${\hat K}_i=\displaystyle \prod_{s=1}^{N_A}
[n^i_s  {\hat a}^{\dagger}_s + (1-n^i_s){\hat a}_s{\hat a}^{\dagger}_s]$,
resulting in:
\begin{eqnarray} \label{K_operator}
\rho_{A}(i,j)=\langle\Psi|{\hat K}_i^{\dagger} {\hat K}_j|\Psi\rangle.
\end{eqnarray}
It is important to note that $\rho_{A}(i,j) \ne 0$ only for states which
have the same number of particles $n^i_A$ in region A, where
$n^i_A=\displaystyle \sum_{s=1}^{N} n^i_s$. 
Thus, in this basis, the reduced density matrix is 
composed of blocks which increases in size with $n^i_A$. Thus for $n^i_A=0$,
the block size is one, for $n^i_A=1$ it is $N_A$, for $n^i_A=2$ it is 
$N_A(N_A-1)/2$, etc.

Thus, the task of calculating the EE of a region A at a given excitation
$|\Psi\rangle$ is equivalent to calculating the eigenvalues of the matrix 
$\rho_{A}(i,j)$. Since the blocks are uncoupled, it is possible
to diagonalize each block with a given number of particles, 
$\rho_{A}^{(\ell)}(i,j)$, independently (where 
$\ell$ denotes the number of particles in region A).
For a state $|\Psi\rangle$, which is the ground state of 
a half-filled ring at $N \rightarrow \infty$, it is possible (in several
different ways) to show that
$S_A = -(1/3) \ln (x) + {\rm Const}$ \cite{amico08}, where $x=N_A/N$. 
For the excited
states the task becomes more difficult, and no simple and general result
for $S_A$ exists \cite{alba09}. Here we will calculate the $S_A$ for 
two-particle excitations.

For the sake of completeness lets first consider single particle state 
entanglement. In this simple case $\rho_{A}(i,j)$ is composed of two
blocks: $\rho_{A}^{(0)}(1,1)$ and $\rho_{A}^{(1)}(i,j)$ 
(where $i,j=1 \ldots N_A$). 
Direct evaluation of Eq. (\ref{K_operator}) for any single-particle state
$|\Psi\rangle=|k\rangle$ results in $\rho_{A}^{(0)}(1,1) = 1-x$, while
$\rho_{A}^{(1)}(i,j) = (1/N) \exp(-\imath p (i-j))$. 
The latter is a Toeplitz matrix,
with one eigenvalue equal to $x$ and $N_A-1$ zero eigenvalues. As expected,
the EE for any single particle eigenstate
$|k\rangle$ is equal to
\begin{eqnarray} \label{single_EE}
S_A=- x \ln(x) - (1-x) \ln (1-x) 
\end{eqnarray}
and does not depend on $|k\rangle$.

A note of caution is in place. Since the eigenvalues for $k$ and $-k$ are
degenerate, any linear combination of $|k\rangle$ and $|-k\rangle$ are
an excited state of Eq. (\ref{clean_hamiltonian}). These linear combinations
have different values of the EE, and therefore, strictly speaking, the
EE for degenerate excited states is ill defined. We circumvent this problem
by introducing a degeneracy breaking magnetic flux $\phi$ into 
Eq. (\ref{clean_hamiltonian}). As long as the degeneracy is broken the EE
of any excited state $|k_1,k_2\rangle$
is well defined and does not depend on $\phi$.

For two-particle states $|k_1,k_2\rangle$, the reduced density matrix
is composed of three blocks: $\rho_{A}^{(0)}(1,1)$, $\rho_{A}^{(1)}(i,j)$
(of size $N_A$) and  $\rho_{A}^2(i,j)$ (size $N_A (N_A-1)/2$).
For the zero particle block:
\begin{multline} \label{rho_0}
\rho_{A}^0(1,1)= \\
\frac{1}{N^2} \displaystyle \sum_{j_1>j_2>N_A}^{N} 
|A(j_1,k_1,j_2,k_2)|^2 =  
(1-x)^2 - y^2,
\end{multline}
where $y =(\sin(\pi(k_2-k_1)x)/(\pi(k_2-k_1)))$.
Thus the eigenvalue of this block is $(1-x)^2 - y^2$.
Using symmetry, one can immediately deduce the eigenvalues of the two-particle
reduced density matrix, without actually diagonalizing the $N_A (N_A-1)/2$
matrix. Since $S_A=S_B$ the contribution to the EE from $\rho_{A}^{(2)}(i,j)$,
must be equal to the contribution from $\rho_{B}^{(0)}(1,1)$. This
infers that  $\rho_{A}^{(2)}(i,j)$ has only one non-zero eigenvalue.
Seeing that region B's length is $N-N_A$, according to Eq. (\ref{rho_0}),
the non-zero eigenvalue of $\rho_{A}^{(2)}(i,j)$ is equal to
$x^2 - y^2$.

The one-particle block density matrix is given by:
\begin{multline} \label{rho_1}
\rho_{A}^{(1)}(i,j)=\frac{1}{N^2} \displaystyle \sum_{j_1>N_A}^{N} 
A^*(j_1,k_1,i,k_2)  A(j_1,k_1,j,k_2)= \\
\frac{e^{-\imath p_1(i-j)}}{N}\bigg((1-x)(1+e^{-\imath (p_2-p_1)(i-j)}) + \\
\frac{1}{\imath N(p_2-p_1)}\big[e^{-\imath (p_2-p_1) i} 
(e^{\imath (p_2-p_1)N_A}-1) \\
-e^{\imath (p_2-p_1) j} (e^{-\imath (p_2-p_1)N_A}-1)\big] \bigg).
\end{multline}
This cumbersome form is substantially simplified when $k_2-k_1$ is
large. In that case the second term in Eq. (\ref{rho_1}) may be neglected
and  the density matrix block has a Toeplitz form
\begin{eqnarray} \label{rho_1_large}
\rho_{A}^{(1)}(i,j)= 
\frac{1-x}{N}\left(e^{-\imath p_1 (i-j)}+e^{-\imath p_2 (i-j)}\right),
\end{eqnarray}
with $N_A-2$ zero eigenvalues, and two degenerate eigenvalues equal to
$x(1-x)$. The second term is negligible also when $x=N_A/N \sim 1/2$, and
$k_2-k_1$ is even, resulting in the same eigenvalues.
We do not have the general solution, nevertheless,
it can be shown numerically that  $\rho_{A}^{(1)}(i,j)$ has no more than two
non-zero eigenvalues, which depend only on the difference $k_2-k_1$.
Moreover, since the sum of all eigenvectors of the density matrix
should be one, the sum of those two eigenvalues should be
$2x(1-x) + 2y^2$.
For the ground state (and excitations for which $k_2-k_1=1$) the two 
eigenvalues are well described by 
$2x(1-x) + (2-1/\pi)y^2$ and $y^2/\pi$.

\begin{figure}
\includegraphics[width=8.5cm,height=!]{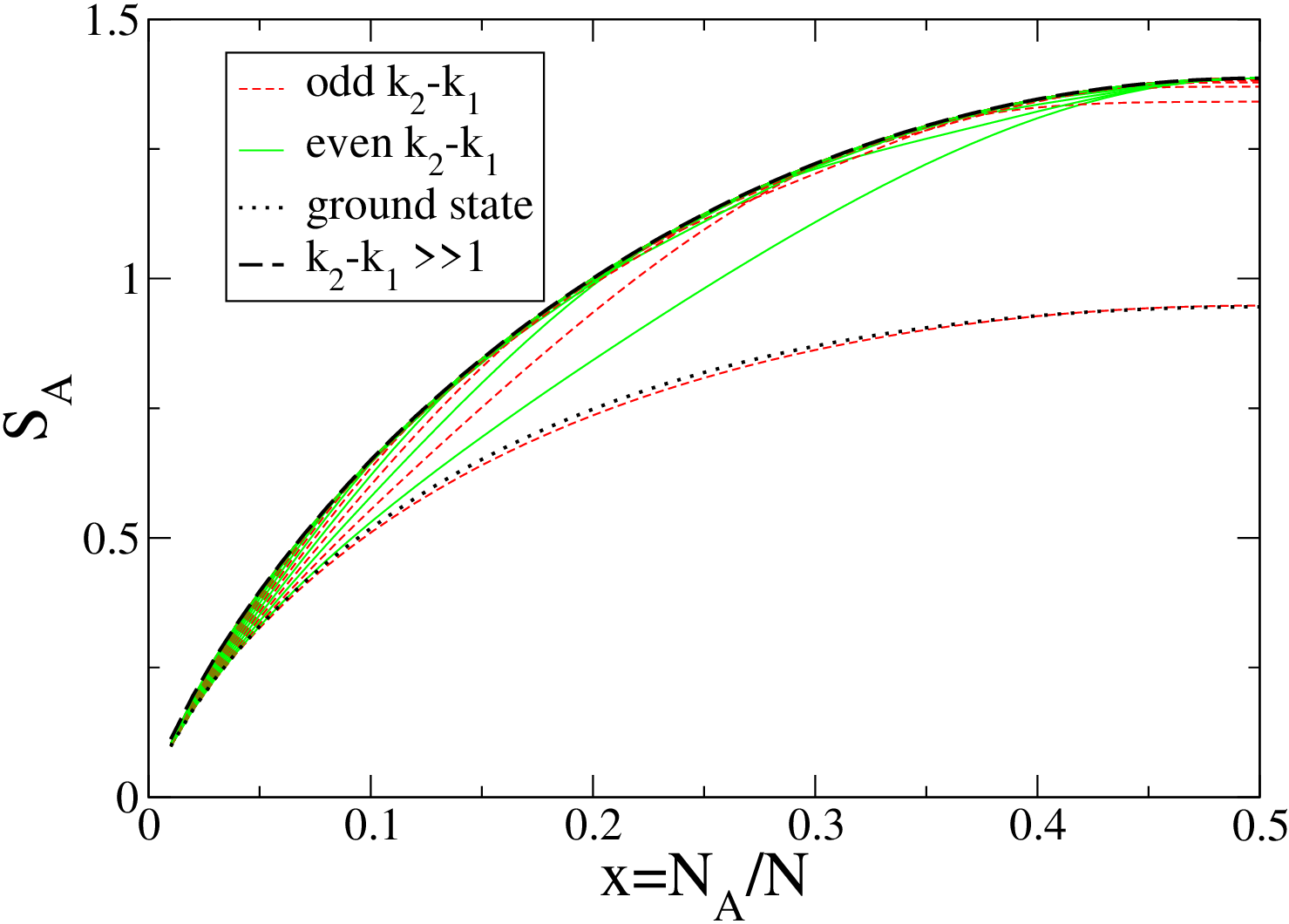}
\caption{\label{fig:clean}
(Color online)
The EE of a two particle state $|k_1,k_2\rangle$ of a clean system as function 
of region A's size $x=N_A/N$ for a system of length $N=1000$.
The Heavy lines correspond to the analytic prediction, (dotted for 
$k_2-k_1=1$, Eq. (\ref{EE_gs}), dashed for $k_2-k_1\gg 1$, 
Eq. (\ref{EE_large_k})). The thin lines pertain to the numerically calculated
EE for all values of $k_2-k_1$ between $1 \ldots 30$, where odd values are
depicted by red curves, and even one by red curves. It is clear that for
$k_2-k_1$ larger than $5$ the numerical curves fit Eq. (\ref{EE_large_k}) 
quite 
well. It is also clear that all even $k_2-k_1$ reach the same EE value
$S_A(x=1/2)=\ln(4)$ once $N_A=N/2$.
}
\end{figure}

Thus, the EE of two-particle states composed from two single-particle 
states of significantly different wave numbers which are 
the majority of the two-particle states, is
\begin{eqnarray} \label{EE_large_k}
S_A(k_2-k_1 \gg 1)= -2\left[(1-x)\ln(1-x)-x\ln(x)\right].
\end{eqnarray}
This is twice the EE of a single particle state (Eq. (\ref{single_EE}). 
Thus, as long as
the two occupied states $|k_1\rangle$ and $|k_2\rangle$ are far enough from
each other, the two-particle EE is just the sum of the EE of each occupied 
state. In the opposite limit 
\begin{multline} \label{EE_gs}
S_A(k_2-k_1=1) \sim -((1-x)^2 - y^2)\ln((1-x)^2 - y^2)\\
-(x^2 - y^2)\ln(x^2 - y^2) -[y^2/\pi]\ln[y^2/\pi] \\
-[2x(1-x)+(2-1/\pi)y^2]\ln[2x(1-x)+(2-1/\pi)y^2].
\end{multline}
The EE curves for other values of $k_2-k_1$ can be calculated numerically
by diagonalizing the $N_A \times N_A$ matrix representing $\rho_{A}^{(1)}$
(the two other eigenvalues for $\rho_{A}^{(0)}$ and $\rho_{A}^{(2)}$ are given
be Eq. (\ref{rho_0})). The results are depicted in Fig. \ref{fig:clean}.
All the values of the EE for any value of $k_2-k_1$ 
fall in between those two limits,
but as can be seen in Fig. \ref{fig:clean}
they quite quickly fall on the $k_2-k_1 \gg 1$
curve. Since there is a large phase space for $k_2-k_1 \gg 1$, a typical 
two particle
excitation corresponds to the EE described in Eq. \ref{EE_large_k}. 

Another interesting behavior that can be gleaned from Fig. \ref{fig:clean}
is that all two-particle states of even $k_2-k_1$ reach the same EE value
at $N_A=N/2$. This stems from the structure of $\rho_{A}^{(1)}(i,j)$
(Eq. (\ref{rho_1})), where the two last terms are multiplied by
$e^{-i(p_2-p_1)N_A}-1 = e^{-i\pi(k_2-k_1)}-1 = 0$ 
(since $p_2-p_1=2 \pi (k_2-k_1)/N$) returning to the density matrix block 
with a Toeplitz form depicted in Eq. (\ref{rho_1_large}), and the 
corresponding two degenerate eigenvalues equal to $1/4$. Since
at $N_A=N/2$, $x=1/2$, and $y =0$, the two other block eigenvalues are also
$1/4$, resulting in $S_A(k_2-k_1={\rm odd},x=1/2) = \ln(4)$. Thus the largest
EE in a two-particle clean ring system is equal to $\ln(4)$.

\section{Interacting Clean Ring}

Incorporating nearest neighbor electron-electron interactions 
into the system,
results in adding an interaction term 
given by
\begin{eqnarray} \label{int_hamiltonian}
H_{\rm int} &=& U
\displaystyle \sum_{j=1}^{N} {\hat a}^{\dagger}_{j}{\hat a}_{j}
{\hat a}^{\dagger}_{j+1}{\hat a}_{j+1}
\end{eqnarray}
to the Hamiltonian, $H$, depicted in Eq. (\ref{clean_hamiltonian}).
In a clean system it is well known that far from half-filling the
system behaves as a Luttinger liquid for any value of $U$ \cite{giamarchi}.
For the ground state EE of a clean system at half-filling  (and $U<2$, i.e., 
a Luttinger liquid) the EE changes only by an overall constant 
\cite{amico08,berkovits12}, while retaining the same logarithmic dependence. 
Thus, we expect that the EE of the two-particle states in a clean system 
will not be essentially affected by
the presence or absence of electron-electron interactions.
Unfortunately, it is not possible to calculate analytically the 
two-particle states of the interacting system. Thus, we must rely on a
numerical solution for the problem.

\begin{figure}
\includegraphics[width=8.5cm,height=!]{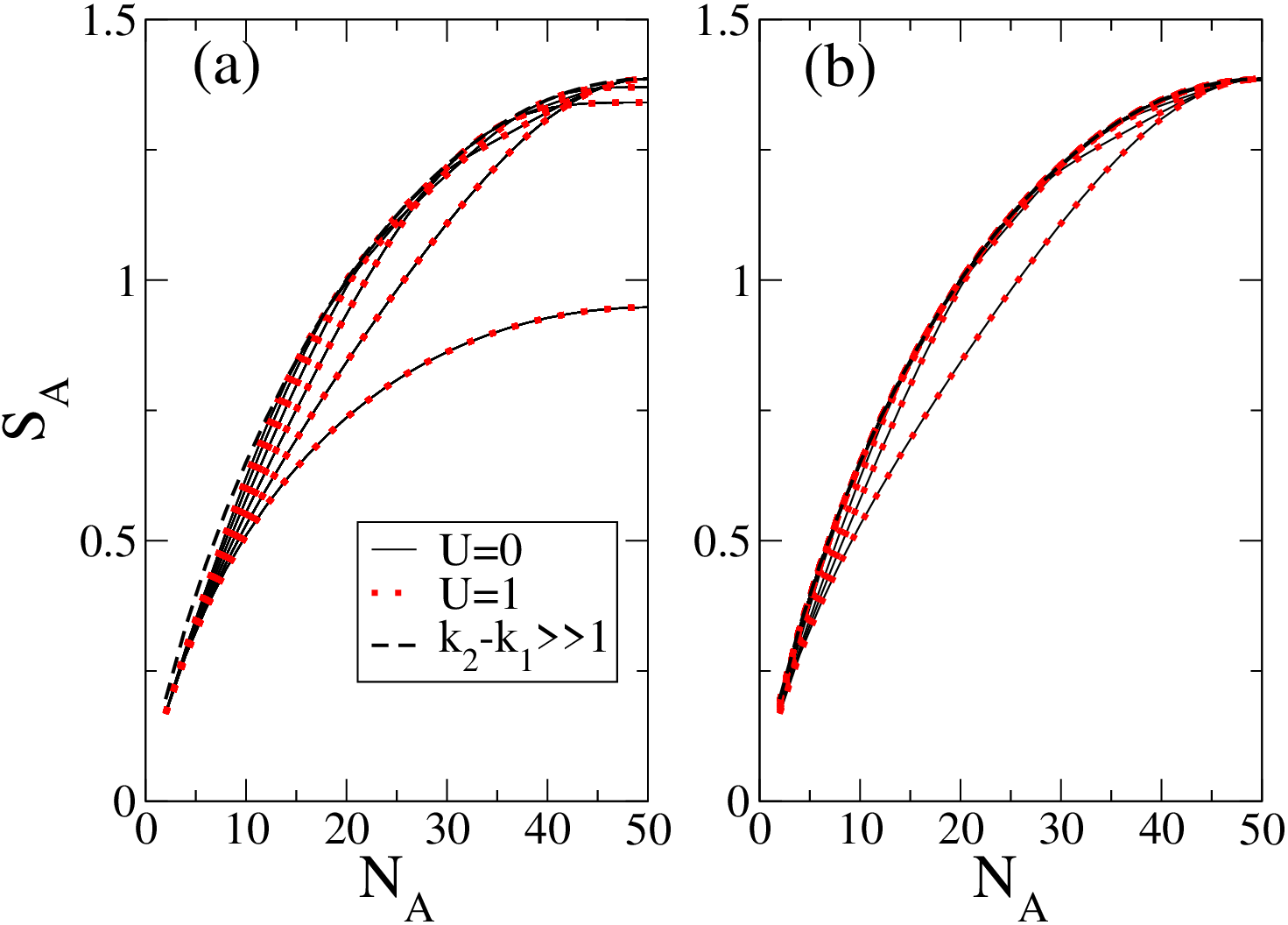}
\caption{\label{fig:int}
(Color online)
The EE of 31 states of a clean system as function 
of region A's size $N_A$ for a system of length $N=100$,
in the absence ($U=0$, black line), or presence ($U=1$, dotted red line) 
of electron-electron interactions.
Panel (a) depicts states in the vicinity of the ground-state 
(the ground state and 1st - 30th excitation), panel (b) shows the excitations
around quarter of the two-particle band (1222th - 1252st excitation).
The Heavy dashed line correspond to the analytic prediction for 
$k_2-k_1\gg 1$, Eq. (\ref{EE_large_k})). 
For a clean system the interaction has no influence on the EE.
}
\end{figure}

Exact diagonalization is used to calculate all
the eigevectors of $H+H_{\rm int}$, represented by a $N(N-1)/2$ matrix.
We have chosen a $100$ site system, resulting in a matrix of size
$4950$. A reduced density matrix $\rho_A$, 
of size $1+N_A+N_A(N_A+1)/2$ is then constructed
and diagonalized for each eigenstate, 
and the EE is calculated using its eigenvectors according
to Eq. (\ref{entropy}). The results are shown in Fig. \ref{fig:int}, 
where the EE of $31$ states around the ground state (i.e. the ground state and
1st - 30th excitation), and at quarter of the two-particle band (1222th -
1252st excitation) are shown. In both cases the EE for the non-interacting
($U=0$) as well as for the interacting ($U=1$) cases are almost equal
(the interacting case is larger by a minute constant (of order $10^{-4}$ 
which can not be resolved at the resolution of the figure). 
As expected
around the ground-state the excitations belong to the low $k_2-k_1$ sector
while for the higher excitations most states corresponds to large values
of $k_2-k_1$, i.e., well described by Eq. (\ref{EE_large_k}).

\section{Interacting Disordered Ring}

When disorder is added to a non-interacting system, all single particle
states become localized. For the many-particle states, the behavior is more 
involved. As long as no interaction is present, the many-particle states remain
localized, both for the ground state \cite{apel82} 
as well as for all the excited
states. Once interaction is introduced, the ground-state as well
as low lying excitations remain localized, while above a critical energy the
many-particle excitations are predicted to delocalize 
\cite{AGKL97,gornyi05,basko06}. 
This transition, termed the many-body or Fock space localization transition,
stems from the interactions coupling excitations with a different number of
electron-hole couples. This type of transition is irrelevant for two particle
systems. Nevertheless, as argued by Shepelynsky and Imry 
\cite{shepelyansky94,imry95}, interaction between the pair of particles
should enhance the two particle localization length, compared to the single
electron localization length, as long as the two-particle level spacing is
significantly smaller than the single electron level spacing, i.e., for higher
excitations.


Can we see any signature of the enhanced two-particle
localization length in the EE 
behavior of the excited states? First we have to understand the influence of
disorder on the EE. In Ref. \onlinecite{berkovits12}  it has been shown
that for the ground-state the EE saturates on the length scale of $\xi$, and
does not continue to grow logarithmic as in a clean system. Thus, the EE of
a disordered system is always lower than the EE of a clean system. One would
expect this feature to hold also for excited states. We check this assumption
by calculating the EE using the excitations of the Hamiltonian given
in Eq. (\ref{clean_hamiltonian}), where the disorder is represented by 
a random on-site energy, $\epsilon_j$ taken from a uniform 
distribution in the range $[-W/2,W/2]$. For $W=3$, single electron states
at the middle of the band are expected to have a localization length 
$\xi \sim 10$ \cite{romer97}, 
while close to the band edge the single electron states are
supposed to be much more strongly localized (Lifshitz tails) \cite{lifshitz64}.

\begin{figure}
\includegraphics[width=8.5cm,height=!]{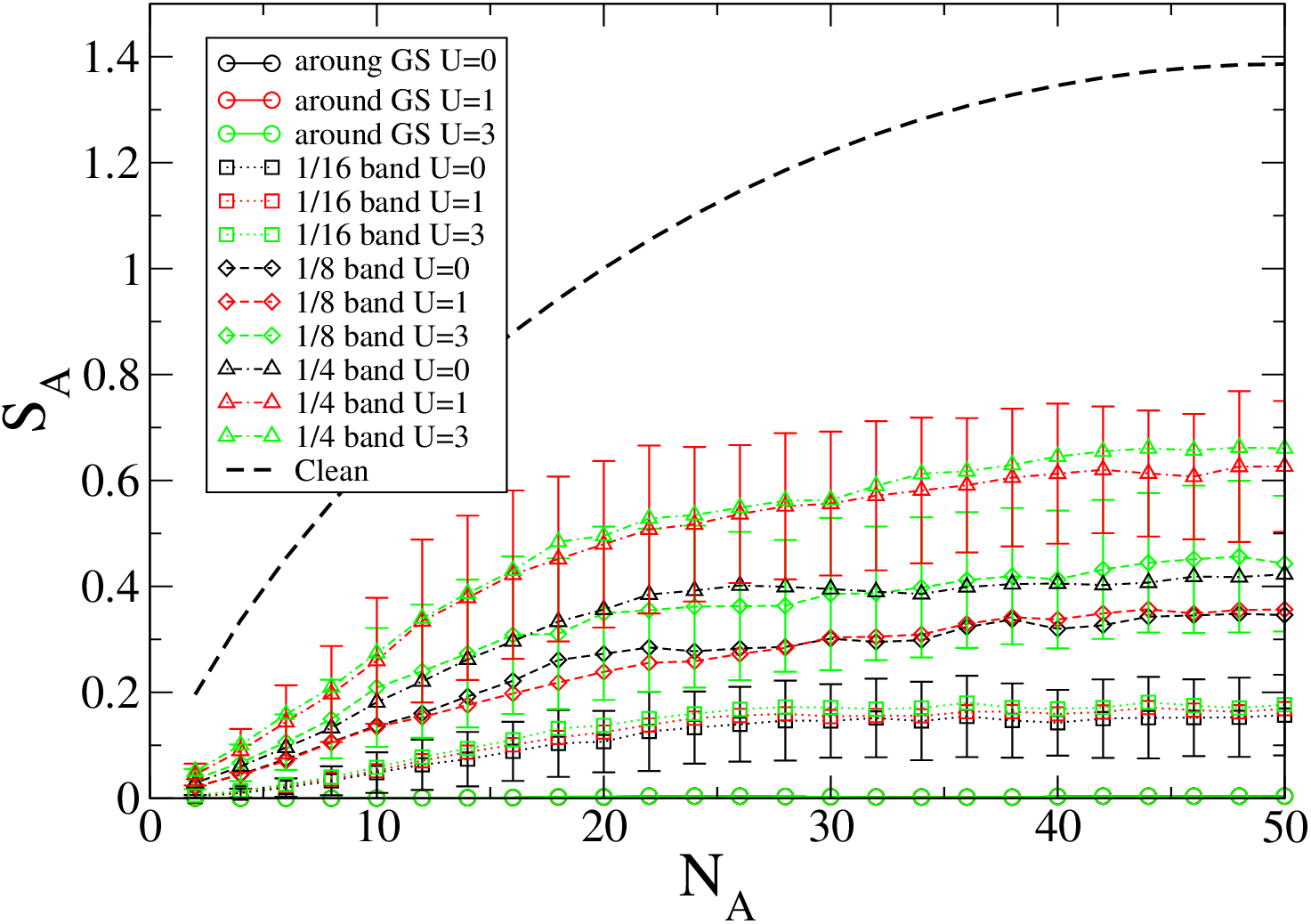}
\caption{\label{fig:dis_strong}
(Color online)
The median EE of an ensemble 
of states for different regions of the band 
collected from 20 different realizations of strong disordered
($W=3$), as function 
of region A's size $N_A$ for a system of length $N=100$.
Continuous lines depict the median EE in the vicinity of the
ground-state (1st - 30th excitation), dotted lines at a$1/16$ of the band
(294th - 324th excitation), dashed lines at a$1/8$ of the band
(603th - 633th excitation), and dot-dashed lines at a$1/4$ of the band
(1222th - 1252th excitation). Black lines correspond to
no electron-electron interactions, while red lines 
indicate the presence 
of moderate electron-electron interactions ($U=1$), and green lines
correspond to stronger interactions ($U=3$).
The heavy dashed line correspond to the maximum EE for a clean system
(Eq. (\ref{EE_large_k})). Error bar represent a range
between the 40th and 60th percentiles. 
}
\end{figure}

The EE is calculated by exact diagonalization for systems of size $N=100$
as described in the previous section. The results are presented in Fig.
\ref{fig:dis_strong}, where the median EE in the vicinity of the
ground-state (1st - 30th excitation), at $1/16$ of the band
(294th - 324th excitation), at $1/8$ of the band
(603th - 633th excitation), and at $1/4$ of the band
(1222th - 1252th excitation). The median EE is taken across the 31 excitations
in each segment and 50 different realizations of disorder.
We calculate the EE in the absence ($U=0$) and presence ($U=1$, $U=3$) of
electron-electron interactions. 
Around the ground state, the interactions do not play a significant role, and
for all cases the EE is strongly suppressed compared to typical values for
a clean system. This is expected, since as mentioned, at the bottom
of the band the states are strongly localized, and therefore the EE is low.

For the non-interacting case, the EE higher in the band are less 
suppressed as the localization length grows. 
The EE for high excitations in the interacting case is 
always larger than for the corresponding non-interacting states.
This is a clear signature for the effect of interactions on localization
of the two-particle states, which become more entangled as interaction
is present, although there is no significant difference between $U=1$ and
$U=3$. It is also clear that for higher excitations (larger localization
length) the enhancement of the EE becomes stronger.

This enhancement could be expected on physical grounds. As has been shown
\cite{shepelyansky94,imry95,frahm95,weinmann95,vonoppen96,jacquod97},
the localization length associated with an interacting two-particle state
is larger than for a non-interacting state with the same disorder. Thus, 
one expects that the EE will also be larger and closer to its clean system
value. 

\begin{figure}
\includegraphics[width=8.5cm,height=!]{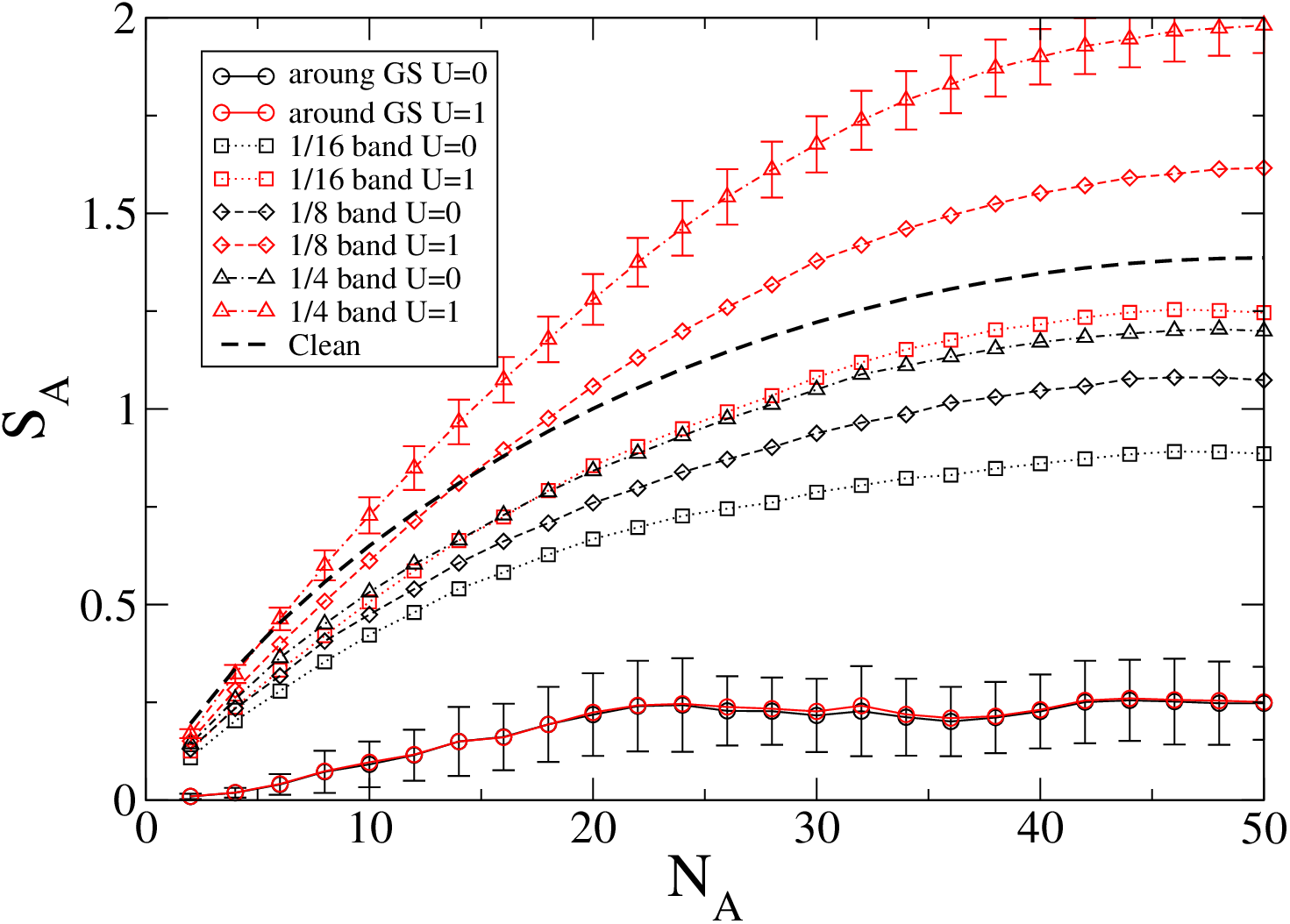}
\caption{\label{fig:dis_weak}
(Color online)
As in Fig. \ref{fig:dis_strong}
for weak disordered
($W=1$). 
For high enough excitation energy, the median EE in the presence
of disorder and interactions exceeds the maximum clean value indicated 
by the heavy dashed line (Eq. (\ref{EE_large_k})). 
}
\end{figure}

We therefore also investigate the case of weaker disorder, for which the
localization length is of order of the system size ($W=1$, $\xi \sim 100$).
As can be seen in Fig. \ref{fig:dis_weak}, for the non-interacting case
a similar pattern to the one observed Fig. \ref{fig:dis_strong} remains,
although the EE is less suppressed by the weaker disorder. As expected,
the enhancement of the EE by interactions is stronger for the weaker disorder.
Surprisingly, above $1/8$ of the band (corresponding to an excitation
energy of $t$), the EE of the disordered interacting system is significantly
larger than the limit for a clean system ($\ln(4)$). Although, extrapolating
from the results presented in  Fig. \ref{fig:dis_strong}, increasing the system
size while keeping the disorder fixed will result in a decrease of the EE
below the clean system values once $L \gg \xi$. The increase above the
clean system excitation EE may stem from the fact that as long as the
two particles are confined within a single localization length the particles
can not avoid each other and spend much time close to each other, 
leading to an enhancement 
of the EE. When the system size is much larger than the localization length, 
the two particles can reside in different regions of the sample, 
and interactions will not
play an important role. However, this hand waving picture requires 
further study. 

At first glance, these results seem to indicate that although
interactions may enhance the EE as long as $L<\xi$, they become
irrelevant for $L \gg \xi$, showing no support for the
many-particle delocalization scenario \cite{AGKL97,gornyi05,basko06}
which should occur for $L \gg \xi$. This interpretation is wrong,
since the many-particle delocalization scenario deals with a constant
density of particles, and delocalization is predicted only when there are
at least a couple of particles in the range of
a single particle
localization length. Thus the observed two-particle EE enhancement when
the two particles are within a distance of $\xi$, as well
as the fact that the enhancement increases significantly when 
the excitation energy, fits nicely with the scenario promoted in Ref. 
\onlinecite{basko06}. Of course, coupling between states
with a different number of electron-hole generation is crucial for the
delocalization scenario, and therefore a full demonstration of the
delocalization transition has to be performed for a finite electron density
system. Nevertheless, the fact that the two-particle behavior 
fits nicely with the delocalization scenario is encouraging.

\section{Conclusions}

The properties of the EE of two particle excited states 
in a one-dimensional ring were studied. 
For a clean system, the EE depends only on
the difference in momentum between the two particles. If the difference
is large the EE corresponds to the EE of two independent single particle
states, i.e., $S_A=- 2[x \ln(x) + (1-x) \ln (1-x)]$. On the other hand,
if the momenta are close, the EE of the two particle state is reduced 
compared to this value. 

One may extrapolate that for $m$ particles on a $N$ site ring, as long
as the density is low ($m/N \ll 1$), the upper limit of the EE is
$S_A=- m[x \ln(x) + (1-x) \ln (1-x)]$, which is also the typical value.
This will be valid if the difference between the momenta of all particles
taking part in a particular many-particle excited state is large. If this
is not the case we expect the EE of the excited state to be lower. Further
investigation of these cases is underway.

We have verified numerically that disorder reduces the EE.
Short range particle interaction leads to an enhancement of the excited state
EE, which become very significant once the localization length is of order of
the system size. For high excitations the median EE of a many
particle interacting excitation is not only above the disordered case, but
exceeds the clean system limit. This may be related to the fact that 
localization forces the two particles to dynamically spend more time in the
vicinity of each other, although this argument merits further study.


\begin{acknowledgments}
Financial support from the Israel Science Foundation (Grant 686/10) is
gratefully acknowledged.
\end{acknowledgments}

\end{document}